\documentclass[USenglish,oneside,twocolumn]{article}
\usepackage[big]{dgruyter_NEW}
\usepackage[utf8]{inputenc}

\cclogo{\includegraphics{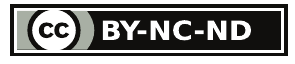}}
\usepackage{subfigure}
\usepackage{balance}
\usepackage{url}
\usepackage{xspace}
\usepackage{verbatim}
\usepackage{epstopdf}
\usepackage{float}

\usepackage{multirow,longtable,colortbl}
\usepackage[font=small,labelfont=bf]{caption}

\newcommand{\descr}[1]{\smallskip\noindent\textbf{#1}}
\newcommand{\name}{{CanaryTrap}\xspace}

\newcommand{\presec}{\vspace{-0.3in}}
\newcommand{\postsec}{\vspace{-0.05in}}
\newcommand{\presub}{\vspace{-0.25in}}
\newcommand{\postsub}{\vspace{-0.1in}}
\newcommand{\presubsub}{\vspace{-0.25in}}
\newcommand{\postsubsub}{\vspace{-0.1in}}

\newcommand{\precaption}{\vspace{-0.05in}}
\newcommand{\postcaption}{\vspace{-0.05in}}
\newcommand{\preequation}{\vspace{-0.15in}}
\newcommand{\postequation}{\vspace{-0.05in}}

\begin{document}

\author*[1]{Shehroze Farooqi}

  \author[2]{Maaz Musa}

  \author[3]{Zubair Shafiq}

  \author[4]{Fareed Zaffar}

  \affil[1]{The University of Iowa, E-mail: shehroze-farooqi@uiowa.edu}

  \affil[2]{The University of Iowa/Lahore University of Management and Sciences, E-mail: maazbin-musa@uiowa.edu}

  \affil[3]{The University of Iowa, E-mail: zubair-shafiq@uiowa.edu}

  \affil[4]{Lahore University of Management and Sciences, E-mail: fareed.zaffar@lums.edu.pk}

\title{\huge \name: Detecting Data Misuse by Third-Party Apps on Online Social Networks} 

\runningtitle{\name: Detecting Data Misuse by Third-Party Apps on Online Social Networks}

\begin{abstract}
{Online social networks support a vibrant ecosystem of third-party apps that get access to personal information of a large number of users.
Despite several recent high-profile incidents, methods to systematically detect data misuse by third-party apps on online social networks are lacking.
We propose \name to detect misuse of data shared with third-party apps.
\name associates a \textit{honeytoken} to a user account and then monitors its unrecognized use via different channels after sharing it with the third-party app.
We design and implement \name to investigate misuse of data shared with third-party apps on Facebook.
Specifically, we share the email address associated with a Facebook account as a honeytoken by installing a third-party app.  
We then monitor the received emails and use Facebook's ad transparency tool to detect any unrecognized use of the shared honeytoken. 
Our deployment of \name to monitor 1,024 Facebook apps has uncovered multiple cases of misuse of data shared with third-party apps on Facebook including ransomware, spam, and targeted advertising.}
\end{abstract}




\maketitle

\presec \presec \section{Introduction} \postsec \postsec

Online social networks such as Facebook and Twitter have millions of third-party applications (or apps) on their platforms \cite{fb9millionApps,onemilliontwitterapps}. 
These third-party apps can potentially get access to billions of accounts including personal information of users who install these apps.
For example, single sign-on (SSO) apps on Facebook typically require access to a user's email address, date of birth, gender, and likes \cite{fbpermissions}. 
Third-party apps with access to personal information of a large number of users have a high potential for misuse.
There have already been several high-profile incidents of data misuse by third-party apps on online social networks \cite{CALeak17GuardianT2,millionFBAccountsAppBought,DigiTrendsSuspendedAppsT5a,googlePlusDataLeak500kT8,googlePlusDataLeak52mT9,appdataleakAWSservers,myPersonalityQuizLeakT6,nameTestDataLeakT7,rankwaveLawsuit,rankwaveFacebook}.
Most notably, a personality quiz app ``thisisyourdigitallife'' harvested data of an estimated 87 million Facebook users that was then used by Cambridge Analytica to create targeted advertising campaigns during the 2016 US presidential election \cite{CALeak17GuardianT2}. 
%
%
%
These incidents have invited scrutiny from regulators.
In July 2019, the FTC imposed sweeping new privacy restrictions on Facebook \cite{ftc5Billion}, including a mandate to suspend third-party apps that do not certify compliance with Facebook's platform policies.
However, the verification of these compliance certifications has proven challenging for Facebook \cite{CambridgeAnalyticaFbNewsRoom,rankwaveLawsuit,rankwaveFacebook}.

There is a lack of methods to systematically detect data misuse by third-party apps.
The main issue is that online social networking platforms lose control over their data once it is retrieved by third-party apps.
These third-party apps can store the retrieved data on their servers from where it can be further transferred to other entities.
Neither users nor online social networks have any visibility on the use of data stored on the servers of third-party apps.
This makes the problem of detecting data misuse extremely challenging since it is hard to track something not under your control.

In this paper, we present \name that uses \textit{honeytokens} to monitor misuse of data shared with third-party apps on online social networks.
A honeytoken refers to a piece of information such as email address or credit card information that can be intentionally leaked or shared to detect its unrecognized (or potentially unauthorized) use  \cite{honeytokenPotNetTerminologies,honeytokenOtherHoneypot,honeytokenCybercriminals}.
\name shares a honeytoken with a third-party app and detects its misuse using different monitoring \textit{channels}. 
For example, if an email address is shared as a honeytoken then received emails act as the channel for detecting unrecognized use of the shared email address.
%

We design and implement \name to investigate misuse of data shared with third-party apps on Facebook.
We share the email address associated with a Facebook account as a honeytoken by installing a third-party app and then monitor the received emails to detect any unrecognized use of the shared email address. 
We conclude that a honeytoken shared with a third-party app has been potentially misused if the sender of a received email cannot be recognized as the third-party app. 
In addition to using emails as a channel to detect data misuse, we also leverage the fact that advertisers on Facebook can use email addresses to target ads to \textit{custom audiences} \cite{customAudienceAdsFB}. 
Specifically, we use Facebook's ad transparency tool ``Why Am I Seeing This?'' to monitor advertisers who have used the shared honeytoken (i.e., email address) to run custom audience ad campaigns on Facebook \cite{fbAdTransparencyHowTo}. 
We conclude that a honeytoken shared with a third-party app has been potentially misused if the advertiser cannot be recognized as the third-party app.

There are two main challenges in scaling \name to monitor a large number of third-party apps that exist on Facebook.
First, we would need to create a Facebook account using a unique email address per-app as a honeytoken.
It is infeasible to create a large number of Facebook accounts because Facebook's anti-abuse systems thwart bulk account registration \cite{fbFakeAccountsActions}. 
To overcome this challenge, we propose an array framework that rotates email addresses associated with a Facebook account while maintaining a one-to-one mapping between shared honeytokens (i.e., email addresses) and third-party apps.
Second, Facebook's anti-abuse systems also limit our ability to frequently rotate email addresses associated with a Facebook account.
Thus, it is infeasible for us to rotate email addresses for each app as we try to scale \name to monitor a large number of third-party apps on Facebook. 
We can install multiple third-party apps on a Facebook account associated with an email address; however, we would not be able to correctly attribute the third-party app responsible for data misuse since the email address was shared with multiple apps.
To overcome this challenge, we propose a matrix framework that uses an $n \times m$ matrix arrangement to install $n m$ apps across a pair of Facebook accounts by rotating $n+m$ email addresses.
This matrix arrangement allows us to attribute the responsible third-party app by establishing a unique two-dimensional mapping for each app.

We deploy array and matrix variants of \name to monitor misuse of data shared with 1,024 Facebook third-party apps. 
We share honeytokens with these apps using array and matrix frameworks and then monitor the misuse through received emails over the duration of more than a year.
By analyzing the emails received on the honeytoken email addresses, the array framework detects 16 apps that share email addresses with unrecognized senders. 
On the other hand, the matrix framework detects 9 out of these 16 apps mainly due to non-deterministic app behavior and implementation issues (which can be readily mitigated in future deployments of \name).
By analyzing Facebook's ad transparency tool, while we are unable to attribute the responsible apps because Facebook does not reveal the email address used by advertisers, we are able to detect unrecognized use of data by 9 advertisers.
%
%

Our further investigation reveals that Facebook does not fully enforce its policies \cite{tos} that require app developers to disclose their data collection and sharing practices as well as respond to data deletion requests by users.
Of the analyzed apps, 6\% apps fail to provide the required privacy policies, 48\% apps do not respond to data deletion requests, and a few apps even continue using user data after confirming data deletion.

\presec \section{Background \& Related Work} \postsec

\subsection{Background} \postsub 
We briefly describe the privacy implications of making user data accessible to third-party apps. 

\descr{Third-party app ecosystem.}
Online social networks provide APIs to enable the development of third-party apps to enhance user experience (e.g., games, entertainment, utilities). 
Popular online social networks such as Facebook and Twitter have millions of third-party apps that are used by hundreds of millions of users \cite{FarooqiOauthAbuse17IMC,appdata,onemilliontwitterapps}. 
For example, Facebook's third-party apps are being used by more than 42,000 of the top million websites \cite{SSOLinkPaper}. 
These third-party apps, with a very large user base in the order of hundreds of millions of users, get restricted access to their users' accounts depending on the permissions granted by the users. 
Third-party apps can request permissions to retrieve user profile information such as their contact information (email address, phone number) and demographics (date of birth, age, gender, sexual orientation) as well as any content they or their friends/followers may have shared (e.g., posts, likes, comments).
While third-party apps are supposed to request the minimal set of permissions needed to implement their functionality, prior research has shown that many third-party apps request more permissions than they actually need \cite{Chia12IsAppSafeWWW}. 
In fact, it is not far fetched to assume that many apps request more permissions than necessary with malicious intent.
In summary, third-party app access to personal information of millions of users opens up possibilities for data misuse. 
Despite its privacy risks, the third-party app ecosystem also benefits users.
For example, the integration of single sign-on (SSO) features into websites expands users' choices for authentication.
Therefore, protecting users from data misuse is important for the growth of this ecosystem.



\begin{figure}[!t]
\centering
\includegraphics[width=1\columnwidth]{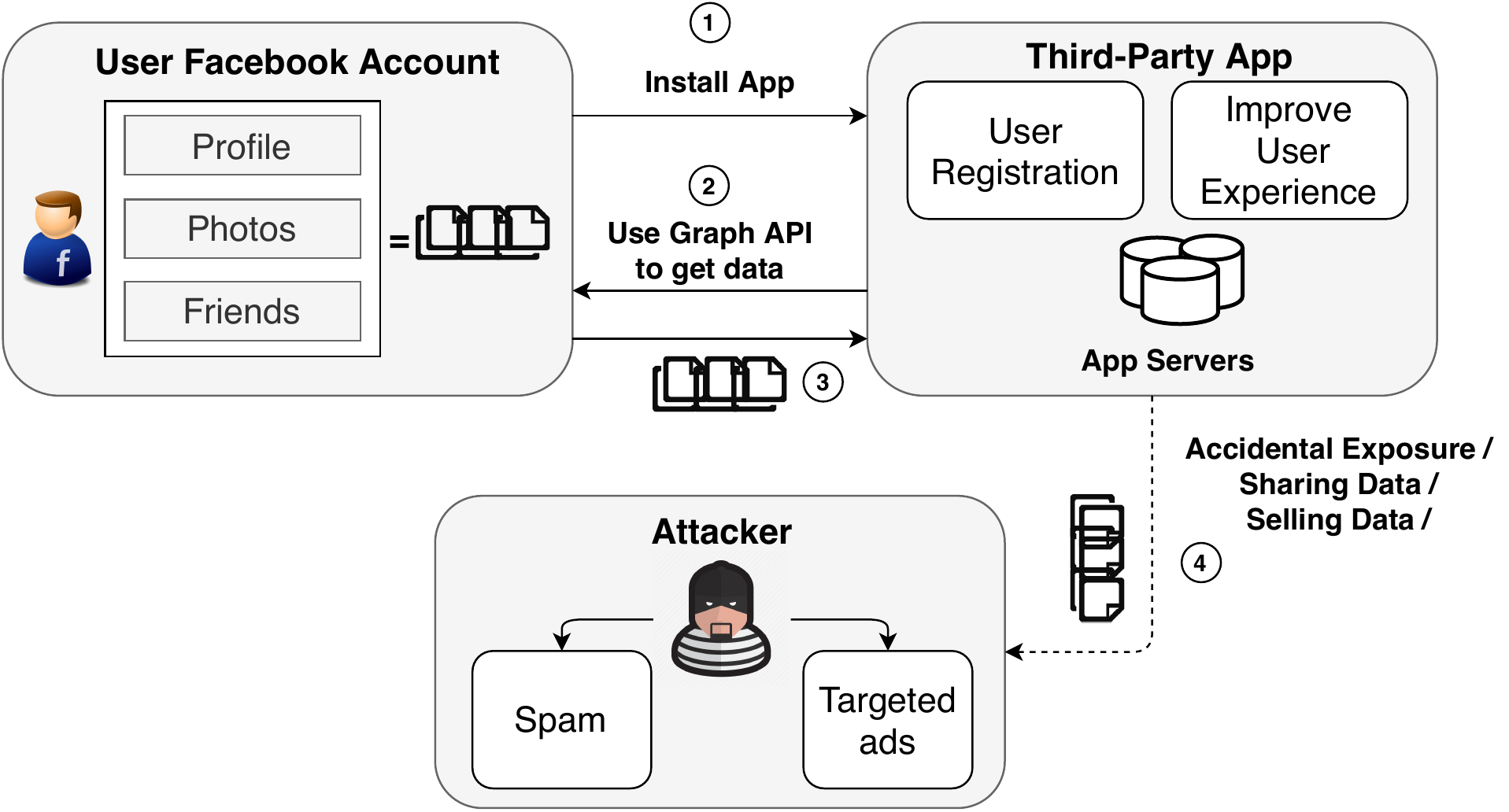}
\precaption
\precaption
\precaption

\caption{Illustration of misuse of data shared with a third-party Facebook app. User data stored on the app server can be leaked to an attacker through unauthorized sale, sharing, or accidental public exposure. The attacker can then misuse this leaked data for various purposes such as marketing through spam or targeted ad campaigns.} 
\label{fig: datamiuseExample}
\postcaption
\postcaption
\postcaption
\postcaption
\end{figure}

\descr{Data leakage/misuse.}
Figure \ref{fig: datamiuseExample} illustrates how a third-party app may engage in data misuse.
Third-party apps use the developer APIs (e.g., Facebook Graph API) to retrieve data of users who install their apps. 
The retrieved data is typically stored on the servers controlled by third-party apps essentially providing them perpetual access to the retrieved data.
These apps can then use the retrieved data to implement their functionality. 
For example, a music streaming app (e.g., Spotify) can leverage the data to suggest relevant music or allow users to share music with their friends or followers.
It is noteworthy that the terms of service (TOS) generally prohibit any use of the retrieved data outside the scope of the third-party app \cite{tos}. 
Hence, third-party apps should not \textit{leak} users' data intentionally (e.g., selling/sharing data to data brokers and advertisers) \cite{rankwaveFacebook} or inadvertently (e.g., accidentally making data publicly available) \cite{appdataleakAWSservers}. 
Leaked data can be misused by an attacker for sending spam emails or targeted ad campaigns. 
We define \text{misuse} as any use of a user's data, which is retrieved by a third-party app, that is outside the scope of the app.

\descr{Role of online social networks in curbing data misuse.}
There have been several recent high-profile incidents of misuse of data shared with third-party apps  \cite{CALeak17GuardianT2,myPersonalityQuizLeakT6,DigiTrendsSuspendedAppsT5a,nameTestDataLeakT7,googlePlusDataLeak500kT8,googlePlusDataLeak52mT9,appdataleakAWSservers,millionFBAccountsAppBought}.
Most notably, a voter-profiling company Cambridge Analytica reportedly used a third-party app ``thisisyourdigitallife'' to harvest personal data (public and private profile information including basic demographics, places visited, interests, and friends) of more than 50 million Facebook users  \cite{CambridgeAnalyticaFiles,CALeak17GuardianT2}.
Data collected through third-party Facebook apps containing user identifiers (user name, email address, user ID) was even reportedly sold on a black market \cite{millionFBAccountsAppBought}.
%
%
%
%
%
These high-profile incidents of data misuse by third-party apps and pressure from regulators has nudged Facebook to audit third-party apps on their platform.
Facebook recently curbed developer access to user data and also started revoking access to dormant APIs \cite{devAccessAPIReduceFB,DigiTrendsSuspendedAppsT5a,fbAppInvestigationLatest}.
%
%
As a part of their recent settlement with the FTC, Facebook now requires developers to annually certify compliance with their TOS \cite{ftc5Billion}.
However, the verification of these certifications has proven challenging for Facebook in the past~\cite{rankwaveFacebook,CambridgeAnalyticaFbNewsRoom}.
Thus, online social networks need to proactively monitor third-party apps on their platform for potential data misuse. 
Unfortunately, online social network operators have skirted their responsibility.
For example, an ex-employee accused Facebook of intentionally limiting their audits of data accessed by third-party apps unless there is negative press or pressure by regulators \cite{nytimeFBManagerPrivacyApps,guardianFBManagerPrivacyApps}.
The callous attitude of online social networks in protecting their users' data from third-party apps further highlights the need for methods that can be independently deployed to detect data misuse by third-party apps.
However, the unavailability of privileged information (e.g., API access logs) and stringent crawling restrictions imposed by online social networks limit the ability of independent researchers and watchdogs to investigate the misuse of data by third-party apps at a large scale.

%

\presub \subsection{Related Work} \postsub

\descr{Detecting data leakage.}
A large body of prior work has focused on detecting leakage of user data on the client-side (e.g., mobile apps, web browsers) to online trackers and advertisers~\cite{Krishnamurthy11privacyleakage,mayer12thirdSP,englehardt15cookiesWWW,starov16contactFormPETS,englehardt18emailTrackingPETS,song2015privacyguardSPSM,ren2016reconMobiSys,renLongitudinalPIINDSS,reyes2018CoppaChildPETS,razaghpanah2018appsPrivacyNDSS,KrishnamurthyPIILeakageWOSN09,HuberAppInspectCOSN13}.
First, some prior work has focused on detecting data leakage in mobile apps through network traffic analysis~\cite{song2015privacyguardSPSM,ren2016reconMobiSys,renLongitudinalPIINDSS,reyes2018CoppaChildPETS,razaghpanah2018appsPrivacyNDSS}. 
For example, Ren et al.~\cite{renLongitudinalPIINDSS} showed that more than 50\% of the 100 most popular mobile apps leak personally identifiable information (PII).
Second, prior work has focused on detecting data leakage in web browsers through cookies, contact forms, or emails~\cite{Krishnamurthy11privacyleakage,mayer12thirdSP,englehardt15cookiesWWW,starov16contactFormPETS,englehardt18emailTrackingPETS}.
For example, Starov et al.~\cite{starov16contactFormPETS} showed that more than 8\% of websites leak users' PII to online trackers through contact forms.
Finally, a subset of prior work on detecting data leakage in web browsers has focused on the leakage of user data by third-party apps on online social networks \cite{KrishnamurthyPIILeakageWOSN09,HuberAppInspectCOSN13}.
Huber et al.~\cite{HuberAppInspectCOSN13} reported more than a hundred examples of PII (e.g., Facebook's user ID and name) being leaked to analytics services by third-party apps.
This line of prior work has two key limitations. 
First, it is limited to detecting data leakage on the client-side -- it cannot detect data leakage on the server-side.
Second, it is limited to detecting data leakage -- it cannot detect potential misuse of the leaked data.
In contrast to prior work on detecting data \textit{leakage}, we focus on the detection of data \textit{misuse} by third-party apps irrespective of whether it is leaked at the client-side or server-side.

\descr{Honeypots.}
Prior work has used honeypots to investigate reputation manipulation in online social networks and account/website compromise.
First, prior work has used honeypots to monitor attackers' interactions with honeypots by leaking credentials \cite{DeBlasioTripwire17IMC,FarooqiOauthAbuse17IMC,OnaolapoHoneyGmail16IMC}.
For example, Onaolapo et al.~\cite{OnaolapoHoneyGmail16IMC} leaked credentials of honey email accounts to blackhat forums and paste sites and monitored activities of attackers who accessed their honey email accounts.
Second, prior work has used honeypots to monitor attacker interactions with honeypots without explicitly handing over their access, such as by purchasing fake followers on Facebook, Instagram, or Twitter \cite{Stringhini13FollowGreenIMC,Cristofaro14PayingForLikesIMC,DeKovenFollowFootstepInsta2018IMC}.
For example, DeKoven et al.~\cite{DeKovenFollowFootstepInsta2018IMC} purchased Instagram followers for honeypot accounts to monitor the activities and infrastructure of spammers.

\descr{Honeytokens.}
Honeytoken is a specialized form of honeypot which is typically a digital piece of information (e.g., email address, credit card information) \cite{honeytokenPotNetTerminologies}.
Honeytokens have been used in prior work to investigate insider threats \cite{spitznerInsiderHT03ACSAC,BowenBaitingDecoy09,kesheeInsiderThreat14IET}, phishing attacks \cite{mcraePhishingHT07HICSS}, and website compromise \cite{DeBlasioTripwire17IMC}.
For example, Spitzner et al. \cite{spitznerInsiderHT03ACSAC} discussed various kinds of honeytokens such as creating a bogus medical record to detect unauthorized access by employees.
More recently, DeBlasio et al. \cite{DeBlasioTripwire17IMC} used email address and password pair as a honeytoken to detect website compromise.
In a similar spirit, we are interested in using honeytokens to detect misuse of data shared with third-party apps in online social networks.

\begin{figure*}[!t]
\centering
\includegraphics[width=2\columnwidth]{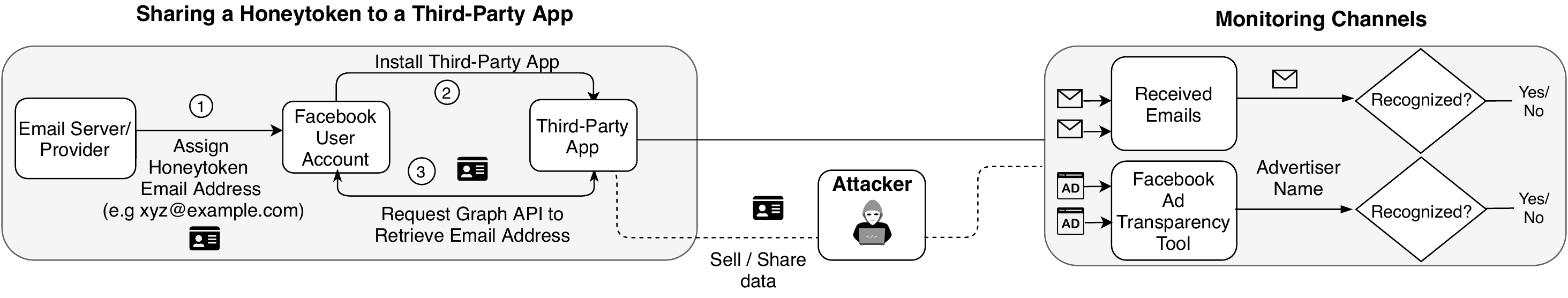}
\precaption
\postcaption
\caption{Overview of \name: (left-to-right) \name associates a honeytoken email address to a Facebook account and then shared it with a third-party app. \name then monitors the misuse of the shared honeytoken through two channels: received emails and Facebook's ad transparency tool.}
\label{fig: proposedapproach}
\postcaption
\postcaption
\postcaption
\end{figure*}

\presec
\section{\name} \postsec

Prior work lacks systematic methods that can be used by online social networks or independent watchdogs to detect misuse of data shared with third-party apps. 
Even though online social networking platforms such as Facebook and Google have unrestricted access to their internal logs, it is challenging for them to detect data misuse by third-party apps because data leaves their platform once a third-party app retrieves it through the APIs. 
Online social networks can look for excessive (or ``anomalous'') API access patterns by third-party apps but they have no visibility into how the data is used by third-party apps once it leaves their premises.
Even though such privileged information (available exclusively to online social networking platforms) can help narrow down the potential misuses, it is desirable to build methods that can be deployed by independent watchdogs without needing cooperation from online social networks.  
Next, we explain the design and implementation of our proposed approach to this end.

\presub
\subsection{Design}\label{ssec:canarytrapdesign}\postsub

\descr{Overview.}
We introduce \name, a \textit{honeytoken} based approach to detect misuse of data shared with third-party apps on online social networks without needing their cooperation. 
Inspired by prior research on honeypots, \name uses a honeytoken to detect misuse of data shared with third-party apps. 
A honeytoken refers to a piece of information such as email address or credit card information that can be intentionally leaked/shared to detect its unrecognized (or potentially unauthorized) use \cite{honeytokenPotNetTerminologies}.
\name shares a honeytoken with third-party apps and detects its misuse using different \textit{channels}.
For example, if an email address is shared as a honeytoken to a third-party app then the received emails act as the channel for detecting unrecognized use of the shared email address.
\name can detect misuse of data shared with a third-party app on any platform (e.g., Facebook, Google) by associating the honeytoken (e.g., email address) to a user account on the platform and then sharing it with the third-party app.
Next, we explain the design of \name and then propose two implementation frameworks that allow its deployment at scale.

\descr{Design Space.}
\name relies on sharing a honeytoken in the data associated with an account that can be used as a bait to trigger misuse. 
There are various attributes in a user account including but not limited to name, email address, date of birth, photos, timeline posts, check-ins, phone number, and address that can be used as honeytokens.
The use of each of the aforementioned attributes as a honeytoken has different pros and cons. 
Below, we discuss the desirable properties in a honeytoken to detect the misuse of data shared with a third-party app.

\begin{enumerate}
    \item \textit{Exploitability}: The honeytoken must provide substantial value and incentives for exploitation.
    
    \item \textit{Soundness}: The honeytoken must allow setting up channels that can be soundly monitored for misuse detection.
    
    \item \textit{Feasibility}: The monitoring channels for the honeytoken must be feasible to set up at scale.
    
    \item \textit{Availability}: The honeytoken must be commonly available in the account information and also commonly requested by third-party apps. 
\end{enumerate}

While we considered various attributes, we ended up selecting \textit{email address} as our honeytoken since it best satisfies the aforementioned properties. 
Next, we justify our choice of email address as the honeytoken.

     
\textit{1)} Email address is \textit{exploitable} by an attacker since it plays a crucial role in the misuse of data due to its uniqueness and also allows the attacker to contact users by sending out emails.
Typically, online services including websites, games, and software require users to associate an email address with their accounts.
Users tend to use the same email address across different accounts \cite{minkusEmailReuse14Pets,PeritoUniqueUsernames11Pets}. 
Hence, email address has emerged as a universal  identifier \cite{emailIsKey1,emailIsKey2,emailIsKey3}.
It also provides data brokers and advertisers the opportunity to unify data acquired on a user from multiple sources to increase its coverage. 
Email address also allows an attacker to contact users by sending them emails.

 \textit{2)} Email address allows us to set up \textit{sound} channels to monitor the misuse of data.
We can set up our own email server or we can partner with an email provider to receive emails on the corresponding email address. 
We can then analyze these emails to detect the misuse of the email address shared with third-party apps.

 \textit{3)} Email address allows us to set up \textit{feasible} channels in a large-scale controlled experiment.
We can acquire a large number of unique email addresses to monitor millions of third-party apps. 
We can either set up our own email server to create any number of email accounts or we can partner with an email provider.

 \textit{4)} Email address is available in the account information of most online platforms.
It is also commonly requested by third-party apps \cite{wangAppPermission11CHIMIT}. 
Moreover, email is one of the few permissions that can be requested by third-party apps on Facebook without requiring a review \cite{loginPermissionsReview}. 


\presub \subsection{Implementation}  \label{ssec:detectabuse} \postsub
Next, we discuss \name's implementation to detect misuse of data shared with third-party apps on Facebook using email address as the honeytoken.

\descr{Sharing a honeytoken with a third-party app.}
We create a fresh Facebook account to share an email address as the honeytoken with a third-party app. 
We can either partner with a major email provider \cite{DeBlasioTripwire17IMC,OnaolapoHoneyGmail16IMC} or set up our own server for email accounts.
We assign an email address to the Facebook account.
We then share the email address with the third-party app by installing the third-party app.
The third-party app is able to access the email address using Facebook's Graph API.

\descr{Monitoring Channels.}
We set up two channels to monitor the misuse of the honeytoken shared with the third-party app monitored by \name. 
First, we use received emails on the email account of the shared honeytoken as a monitoring channel. 
We conclude that a honeytoken shared with a third-party app has been potentially misused if the sender of a received email cannot be recognized as the third-party app. 
Second, we use Facebook's ad transparency tool ``Why Am I Seeing This?'' \cite{fbAdTransparencyHowTo,fbAdTransparencyCustomAudience} as a monitoring channel. 
We monitor whether advertisers use the shared email address to target ads to Facebook \textit{custom audiences}   \cite{customAudienceAdsFB}.
%
%
We conclude that a honeytoken shared with a third-party app has been potentially misused if the advertiser cannot be recognized as the third-party app.
We explain this process in detail below.


\textit{1) Detecting data misuse using received emails.}
Given a received email and a third-party app (app's name and domain name) as input, we use \textit{keyword matching} to determine whether the sender of the email is the input third-party app. 
To match an email with a third-party app, we generate \textit{keywords} using the app's name and its host website's domain name. 
For an app's name, we start with its full name and create tokens of the name if it is composed of multiple words.
For example, if the app name is ``test application'' we generate  ``test application'', ``test'', and ``application'' as keywords.
For the domain name, we use the complete domain name and also create tokens of domain levels, excluding TLDs (e.g., .com, .co, .co.uk).  
For example, if the domain name is ``subdomain.example.com'' we create ``subdomain.example.com'', ``subdomain'', and ``example'' as keywords.
We then search these keywords in the header of the received email.
Specifically, we search for them in ``from'', ``reply-to'', ``message-id'', and ``subject'' fields.
If any of the keywords is found in any of these fields, we label the email as \textsf{recognized}; otherwise, we label the email as \textsf{unrecognized}.
We also call the sender of an \textsf{unrecognized} email as \textit{unrecognized sender}.
The unrecognized sender can be a partner website or an external service used by the app to send emails. 
Therefore, we manually analyze the content of \textsf{unrecognized} emails to further determine data misuse.




\textit{2) Detecting data misuse using Facebook's ad transparency tool.}
We identify the names of advertisers who upload our shared honeytoken by crawling the advertiser information provided by Facebook's ad transparency tool \cite{fbAdTransparencyHowTo,fbAdTransparencyCustomAudience}.
Specifically, Facebook's ad transparency tool provides a list of advertisers who uploaded a list with an email address associated with a Facebook account \cite{VenkatadriPPITargettingDataBroker18SP}.
For each of the listed advertisers, we extract the advertiser's name and its domain (if available) listed on their Facebook page.
We then match the advertiser's name and its domain against the app's keywords generated using the aforementioned keyword matching process. 
If any of the keywords is matched with the name of the advertiser, we label the advertiser as \textsf{recognized}; otherwise, we label the advertiser as \textsf{unrecognized}.
We conclude that data has been potentially misused by an \textsf{unrecognized} advertiser because Facebook's ad transparency tool provides the evidence that the email address shared with the third-party app has been uploaded by the \textsf{unrecognized} advertiser with Facebook.

\descr{Summary.}
Figure \ref{fig: proposedapproach} provides an overview of \name to detect misuse of data shared with third-party apps on Facebook.
\name uses a Facebook account to share an email address as a honeytoken with a third-party app.
Misuse is detected using two channels: by analyzing the received emails and the advertisers listed by Facebook's ad transparency tool.
%

\presub \subsection{Deployment} \label{ssec:deployment} \postsub
Since online platforms support a large number of third-party apps, it is important that \name's deployment can scale.
We face two challenges in scaling \name's deployment on Facebook.

First, \name requires creating a Facebook account to share a unique email address as a honeytoken with a third-party app.
Since Facebook's anti-abuse systems thwart automated/manual bulk account registration \cite{fbFakeAccountsActions}, it is infeasible to create a large number of Facebook accounts to monitor a large number of apps.
To overcome this challenge, we propose an \textit{array framework} that rotates multiple email addresses associated with a Facebook account to maintain  one-to-one mapping between the shared honeytokens (i.e., email addresses) and third-party apps.
Note that we can practically create as many email accounts as necessary by setting up our own email server as we discuss later in Section \ref{ssec:experimentalsetup}.

Second, Facebook's anti-abuse systems also limit our ability to frequently rotate the email address associated with a Facebook account.
Thus, it is infeasible for us to rotate email addresses for each app as we try to scale \name to monitor a large number of third-party apps on Facebook. 
As an alternate, we can share the same honeytoken to multiple third-party apps using a single Facebook account.
However, in case a third-party app misuses our data, it would not be possible to identify the responsible app since the email address was shared with multiple apps.
To overcome this challenge, we propose a \textit{matrix} arrangement to install $n \times m$ apps across a pair of Facebook accounts by rotating $n+m$ email addresses.
This matrix arrangement enables us to  attribute the responsible third-party app by establishing a unique two-dimensional mapping for each app.

The matrix framework is more scalable than the array framework. 
More specifically, the matrix framework allows us to reuse a honeytoken for monitoring multiple third-party apps. 
It is able to provide a unique app-to-honeytoken mapping by sharing two honeytokens with each app in a two-dimensional arrangement. 
It allows \name to monitor $N$ apps using as little as $2\sqrt{N}$ honeytokens, instead of needing $N$ honeytokens in the array framework implementation.

Next, we detail the deployment of \name using both array and matrix frameworks. 


\presubsub \subsubsection{Array Framework} \postsubsub
In the array framework, we install all third-party apps on a single Facebook account such that a unique honeytoken (email address) is shared with each app.
To this end, we create a single Facebook account, $A$, and set up different honeytoken email addresses for it, $HT_A$. 

\preequation\vspace{-0.05in}
\[F = \{A\}\ , HT_A = \{e_{1},e_{2},...,e_{n}\}\ , App = \{a_1,a_2,...,a_n\}\]
We start by associating an email address $e_i$ from the $HT_A$ to the Facebook account $A$ and install a new app $a_i$. 
We then uninstall the app $a_i$ and remove the email address $e_{i}$. 
This ensures that the app $a_i$ is only shared a single honeytoken i.e., $e_{i}$.
We then repeat this process for all of the apps in $App$ such that the app $a_i$ is shared with the honeytoken $e_{i}$.

Since only the third-party app $a_i$ knows the email address $e_i$, any email received on $e_i$ is attributed to the third-party app $a_i$.
Let's assume that $e_i$ receives $k$ emails.
We denote emails received by $e_i$ as $Mail_{{e_i}} = \{mail_1,mail_2,...,mail_k\}$.
%
We then input each email from $Mail_{e_i}$ and third-party app $a_i$ to the matching process explained in Section \ref{ssec:detectabuse} to check whether the email is labeled as \textsf{recognized} or \textsf{unrecognized}.
We say that an app $a_i$ is responsible for the unrecognized use of honeytoken shared with it if one or more emails from $Mail_{e_i}$ are labeled as \textsf{unrecognized}.

\presubsub \subsubsection{Matrix Framework} \postsubsub
The matrix framework allows us to  attribute the third-party app responsible for the misuse while sharing a honeytoken to multiple apps.
To monitor $N$ apps using matrix framework, we require $n$ and $m$ honeytokens such that $N<=n \times m$. 
If $N$ is a perfect square then $n=m=\sqrt{N}$.
If $N$ is not a perfect square then $n \ge \sqrt{n}$ and  $m \ge \sqrt{n}$ and $N<n \times m$.

\begin{figure}[!t]
\centering
\includegraphics[width=0.7\columnwidth]{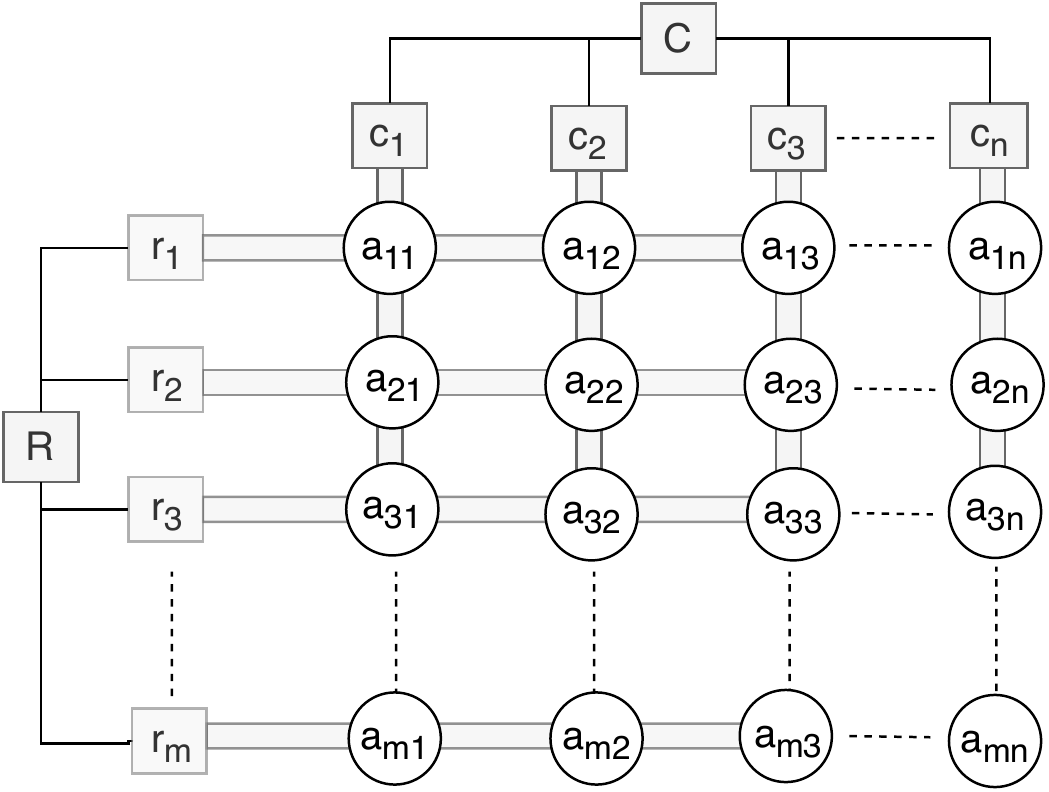}
\precaption
\caption{Illustration of the matrix arrangement used by \name. Two sets of email addresses ($\{r_{1}, ..., r_{m}\}$ and $\{c_{1}, ..., c_{n}\}$) are associated with two Facebook accounts (R [row] and C [column]). While each honeytoken is shared with multiple apps in the corresponding row or column, an app is shared two honeytokens (one from row and one from column) to create a unique two-dimensional mapping.}
\label{fig: LinMat}
\postcaption
\postcaption
\postcaption
\postcaption
\end{figure}

To implement \name using a 2-dimensional matrix framework, we create two Facebook accounts, \textit{R} and \textit{C}. 
Each account is associated with a set of unique email addresses that will act as our honeytokens.
We share two unique email addresses with each third-party app, one email address associated with the Facebook account \textit{R} (row) and one email address associated with the Facebook account \textit{C} (column).
Figure \ref{fig: LinMat} shows the deployment of \name using two Facebook accounts associated with \textit{n} and \textit{m} email addresses. 
This matrix arrangement allows us to monitor data misuse by $n \times m$ third-party apps using $n+m$ honeytokens. 

\preequation
\[F = \{R,C\}\] 
\[HT_R = \{r_{1},r_{2},...,r_{m}\}\ ~~~ HT_C = \{c_{1},c_{2},...,c_{n}\}\]
\precaption
\[ App =
  \left[ {\begin{array}{ccccc}
   a_{1,1} &  .. & .. & a_{1,n}\\
    . & .  & .  & .\\
   a_{m,1} &  .. & .. & a_{m,n}\\
  \end{array} } \right]
\]
\postequation

We start setting up the matrix framework with Facebook account \textit{R} and associate one email address at a time with the account and install all the apps assigned to this honeytoken. 
This corresponds to installing all the apps in a given row shown in Figure \ref{fig: LinMat}. 
We then remove all of the installed apps and repeat this process for all email addresses in $HT_R$.
We repeat the process for Facebook account \textit{C} and the email addresses in $HT_C$.
Once this process is completed for account \textit{C}, each app has been shared two honeytokens, one from account \textit{R} and one from account \textit{C}. 
Specifically, each app $a_{i,j}$ from $App$ has been shared exactly two email addresses $r_{i}$ and $c_{j}$. 


To correctly attribute an email to a third-party app, we use the following insight.
Any email received on email account $r_{1}$ can belong to any of \textit{n} apps $a_{1,j}$ where $j \in \{1,2,...,n\}$ corresponding to its row in the matrix. 
Similarly, any email received on email account $c_{1}$ can belong to any of \textit{m} apps $a_{i,1}$ where $i \in \{1,2,...,m\}$ corresponding to its column in the matrix. 
Recall that each third-party app $a_{i,j}$ is installed on two accounts \textit{R} and \textit{C}, respectively, and is shared exactly two honeytokens $r_{i}$ and $c_{j}$. 
Hence, if a third-party app $a_{i,j}$  sends out any email to their users, such an email would be received by one honeytoken $r_{i}$ in $HT_R$ and one honeytoken $c_{j}$ in $HT_C$. 
Specifically, we collect all the emails received by the pair of email addresses $r_{i}$ and $c_{j}$ assigned to an app $a_{i,j}$.
If the sender of an email received at $r_{i}$ is the same as the sender for an email at $c_{j}$, we attribute the email to the third-party app $a_{i,j}$.
We then input each of these attributed emails and third-party app $a_{i,j}$ to the process explained in Section \ref{ssec:detectabuse} to check whether the email is labeled as \textsf{recognized} or \textsf{unrecognized}.

Note that this attribution strategy is susceptible to mistakes if a pair of apps send emails from the same email address. 
To understand this issue, consider the following scenario. 
Let's assume that two apps $a_{1,2}$ (that has been shared honeytokens $r_1$ and $c_2$) and $a_{3,1}$ (that has been shared honeytokens $r_3$ and $c_1$) send emails from the same email address. 
The emails from $a_{1,2}$ and $a_{3,1}$ received at $r_1$ and $c_1$, respectively, will be incorrectly attributed to $a_{1,1}$.
Similarly, the emails from $a_{1,2}$ and $a_{3,1}$ received at $c_2$ and $r_3$, respectively, will be incorrectly attributed to $a_{3,2}$.
These emails are ``conflicting'' because they are attributed twice, i.e., once when they are correctly attributed to their respective app and once when they are incorrectly attributed to another app.
We exclude these conflicting emails to avoid mistakes which can result in falsely labeling an email as \textsf{unrecognized}.


\presec 
\section{Results} 
\postsec
In this section, we first explain our experiments to deploy \name to monitor 1,024 third-party apps on Facebook.
We then investigate cases of data misuse detected by \name by analyzing the received emails and Facebook’s ad transparency tool.
Finally, we evaluate the effectiveness of \name's matrix framework in detecting data misuse.

\presub \subsection{Experimental Setup}  \label{ssec:experimentalsetup}
\postsub

\noindent \textbf{Identifying third-party apps.}
%
Since Facebook does not provide an index of third-party apps,\footnote{Facebook  discontinued the app center  \cite{HuberAppInspectCOSN13}. It now only lists game apps.} 
we use a list of Facebook apps from prior research~\cite{SSOLinkPaper} that was compiled by crawling the web.
This list contains 43,332 \textit{host websites} that integrate Facebook apps (e.g., SSO, social plugins). 
By re-crawling these 43,332 host websites, we find active Facebook apps on 25,800 websites that request email address. 
We randomly select 1,024 of these host websites that integrate Facebook apps to deploy CanaryTrap for monitoring data misuse.
Our random selection includes different services such as music streaming and news sites.

\noindent \textbf{Setting up email honeytokens.}
The array framework requires sharing one email honeytoken per app.
Thus, we need 1,024 honeytokens to monitor 1,024 third-party apps using the array framework.
Matrix framework requires sharing $2\sqrt{mn}$ honeytokens to monitor $mn$ apps. 
Thus, we need 64 honeytokens to monitor 1,024 third-party apps using the matrix framework.
To this end, we set up a .com email server and use a list of popular names  \cite{SurnamesGithub} to create email accounts using the firstname-lastname@example.com template.
For example, if we have a first name ``john'' and last name ``doe'', we generate an email account with email address ``john-doe@example.com''.

\noindent \textbf{Setting up Facebook accounts.}
We register three fresh Facebook accounts in total: one Facebook account $A$ for the array framework and two Facebook accounts $R$ and $C$ for the matrix framework. 
We set the privacy settings of these accounts such that their personal information, including email addresses, remains private to everyone except for the installed apps.

\begin{figure}[!t]
\centering
\includegraphics[width=0.85\columnwidth]{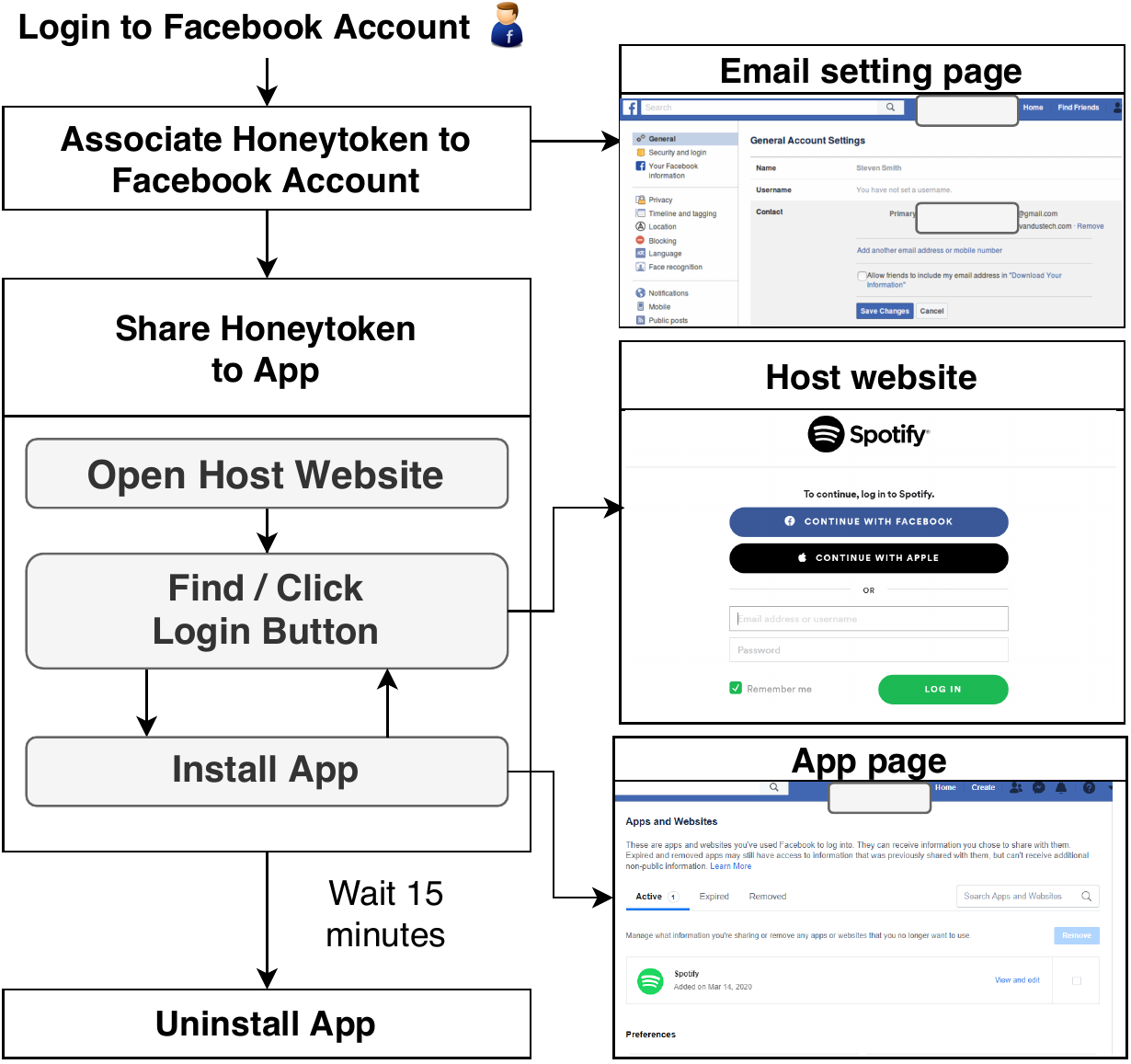}
\precaption
\caption{Automated workflow of our CanaryTrap implementation to associate a Facebook account to honeytokens that are then shared with third-party apps using the array and matrix frameworks.}
\label{fig: crawlerWorkflow}
\postcaption
\postcaption
\postcaption
\postcaption
\end{figure}

\noindent \textbf{Workflow of sharing honeytokens with third-party apps.}
We use Selenium to automate the process of (1) \textit{associating honeytokens to Facebook accounts} and then (2) \textit{sharing honeytokens with Facebook apps}.
Our implementation ensures that honeytokens are appropriately rotated in the Facebook accounts and shared with the aforementioned 1,024 Facebook apps using the array and matrix frameworks.
Figure \ref{fig: crawlerWorkflow} shows the workflow of our implementation.

\textit{Associating honeytokens to a Facebook account:}
To this end, we open the Facebook account's email settings page \cite{fbGeneralSettingsTab} to add the honeytoken email address to the account and select it as the primary email address.
Note that a Facebook account can have multiple associated email addresses but only the primary email address is accessible to an app. 
Both array and matrix framework implementations require rotating email addresses in Facebook accounts.
Array framework requires rotating email addresses more frequently than the matrix framework since the array framework has a one-to-one mapping between a honeytoken and an app. 
More frequent rotation of email addresses is challenging because Facebook's anti-abuse systems rate limit the addition of email addresses to the account.
In comparison to the array framework, the matrix framework is more amenable to operating under these rate limits.

\textit{Sharing honeytokens with Facebook apps:}
To this end, we open the host website to install the app.
The diverse user interfaces of host websites make it challenging to automate the process of installing the Facebook app. 
%
By surveying a sample of host websites, we design templates of regular expressions such as \texttt{contains(@href, `facebook')} and \texttt{contains(@class, `login-fb')} to find Facebook login buttons from DOM elements on the host websites.
We then look for Facebook login buttons using a list of candidate URLs for each host website.
This list includes the host website's landing page, host website's URL that has a Facebook app identified by prior work~\cite{SSOLinkPaper}, and URLs returned from Google search by issuing a query of login/sign up pages for the host website.
We click on the buttons that match our templates of regular expressions on each candidate URL page until we find a Facebook login button that redirects to Facebook to install the app.
After the app has been successfully installed, we are redirected back to the host website. 
Upon installation, the third-party app can  request our email address using Facebook's Graph API.
We allow a grace period of 15 minutes to the app for accessing our email address before uninstalling it and installing the next app. 
We need to uninstall the app before rotating the email address to avoid exposing the new email address to the last app.
We acknowledge that we might miss potential misuse that will occur as a result of additional interactions (e.g., completing user registration) with the host website or if the app requests our email address after it is uninstalled.

\noindent \textbf{Deployment of CanaryTrap.}
Our implementation of CanaryTrap associates Facebook accounts to honeytokens that are then shared with 1,024 Facebook apps using the array and matrix frameworks around the same time in December 2018.
We manually eyeball screenshots of a sample of 100 apps 
to estimate the success of installing Facebook apps and sharing honeytokens with these apps.
%
Specifically, we analyze two types of screenshots after the app is installed into the Facebook account.
The first screenshot captures the Facebook account's apps settings page to validate whether or not the app has been installed.
We miss the installation of 7\% of apps.
The second screenshot captures the current state of the host website after the app has been installed.
Even if apps are successfully installed as indicated by the first screenshot, we find that honeytokens may not have been shared with some apps due to multiple reasons. 
First, we observe that the host website of some apps after installation gives us various website errors such as page not found (404) or service unavailable (503). 
The fraction of such errors is 6\%. 
%
Second, we observe that host websites of some apps require us to complete additional steps (e.g., additional demographic information) resulting in incomplete account registration on the host website.
We estimate that the fraction of incomplete account registration is 31\%.
%
For the remaining apps, we can confidently say that the apps have been installed and honeytokens have been shared successfully. 



\presub  \subsection{Misuse Detection using Emails} \postsub 
We deploy \name's array framework to monitor 1,024 apps.
We monitor them using email as the monitoring channel for more than a year to detect data misuse. 
We receive 12,704 emails on the email accounts associated with the honeytokens shared with 332 out of 1,024 monitored apps.  
Using the keyword matching process explained in Section \ref{ssec:detectabuse}, 12,282 of the emails are labeled as \textsf{recognized} while the remaining 422 emails are labeled as \textsf{unrecognized}.
These 12,282 \textsf{recognized} and 422 \textsf{unrecognized} emails are received on honeytokens shared with 327 and 20 Facebook apps, respectively. 

Next, we evaluate the accuracy of our keyword matching process to label emails as \textsf{recognized} and \textsf{unrecognized}.
We then characterize the \textsf{unrecognized} use of our honeytokens shared with the Facebook apps. 
Finally, we discuss the potential impact of the Facebook apps that are responsible for the unrecognized use of honeytokens shared with them.

\presub \vspace{-0.05 in}
\subsubsection{Evaluating Recognized and Unrecognized Emails} 
\postsub \postsub
We manually analyze the 422 \textsf{unrecognized} emails and a 1\% sample of \textsf{recognized} emails. 
Our results show that we are able to match emails with apps with high precision.
Specifically, none of the \textsf{recognized} emails are incorrectly labeled while only 69 \textsf{unrecognized} emails shared with 4 apps are incorrectly labeled.
The senders of these incorrectly labeled emails are \textsf{unrecognized} because the email addresses of these senders are from a different domain which does not match with the host website. 
In the future, we can improve the accuracy of the matching process by matching domain registrant information of these domains with the host website since some apps use a different domain as their email server.
For example, the host website \texttt{rajkamalprakashan.com} of an app and their email server domain \texttt{rajkamalbooks.in} that sends \textsf{unrecognized} emails both have the same domain registrant information.
We can also try matching the destination URL of these domains with the host website since some of these domains redirect to the host website.
For example, the domain \texttt{getscoop.com} that sends \textsf{unrecognized} emails redirects to their app's host website \texttt{ebooks.gramedia.com}.
As an alternative to matching the email sender, our existing keyword matching process could also search for keywords in the email's body to reduce these incorrectly labeled emails.
However, we decide not to search in the email's body because it may result in incorrect labeling of a large number of \textsf{unrecognized} emails as \textsf{recognized} due to the occurrence of commonly used words (e.g., application, game) in the email's content.



\begin{figure*}[!t]
\centering
\includegraphics[width=2\columnwidth]{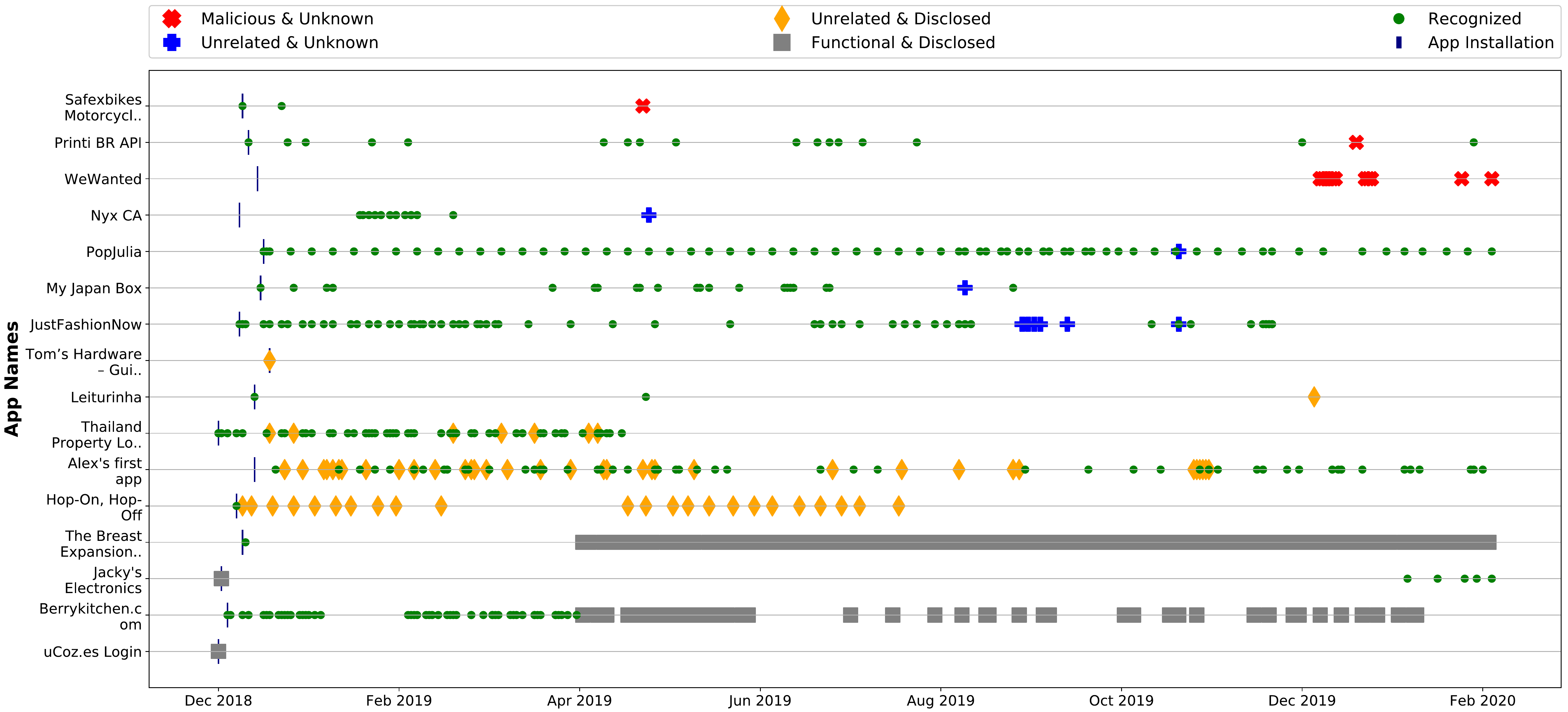}
\caption{Our deployment of \name's array framework on Facebook using emails as a monitoring channel for more than a year uncovers 16 cases of unrecognized use of data by apps. For example, the email accounts of the honeytokens shared with ``Safexbikes Motorcycle Superstore'' app and ``Printi BR API'' app each receive an \textsf{unrecognized} email that asks for ransom through bitcoin falsely claiming that the victim's browsing history has been compromised.}
\label{fig: datamisuse}
\postcaption
\postcaption
\postcaption
\postcaption

\end{figure*}


\presub\subsubsection{Characterizing Unrecognized Emails}\postsub
We characterize the unrecognized use of the honeytoken shared with 16 apps by (1) uncovering the relationship between the unrecognized sender and the app through a disclosure test and (2) analyzing the content of the \textsf{unrecognized} emails. 
This characterization helps us devise a taxonomy that determines the severity of the potential misuse of the honeytoken.

\descr{Disclosure test.}
Our disclosure test checks whether the relationship between the unrecognized sender of an \textsf{unrecognized} email and app is \textit{disclosed} or \textit{unknown}.
%
This test includes any disclosure of the relationship between the unrecognized sender and the app on the host website's landing page, terms, privacy policy, and social media pages. 
If we find any disclosure through this test, we say that the relationship between the app and the unrecognized sender is \textit{disclosed}, otherwise, it is \textit{unknown}.
For example, we receive emails that contain property listings from an unrecognized sender \texttt{dotpropertygroup.com} on the honeytoken shared with the ``Thailand Property Login'' app.
Through the disclosure test, we identify a disclosed relationship between the ``Thailand Property Login'' app and \texttt{dotpropertygroup.com} on the landing page of the app's host website.
In case of a disclosed relationship, the unrecognized senders are typically a partner/affiliate website, an external service (e.g., user authentication service), or a company that acquired the Facebook app.
In case of an unknown relationship, the unrecognized senders potentially get access to the user's data through breaches or leakages on the app's servers \cite{appdataleakAWSservers} or secret data sharing deals between apps and unrecognized entities.
We acknowledge that an app is not culpable if it legitimately shares an email address with an external service (e.g., a bulk email service) and the external service accidentally or deliberately leaks the email address to an unrecognized sender. 
Out of the 16 apps, we find that 9 apps have a disclosed relationship with the unrecognized senders while the remaining 7 apps have an unknown relationship.

\descr{Content analysis.}
We manually analyze the content of the 353 \textsf{unrecognized} emails received from the unknown and disclosed senders and label them to either (1) ``Malicious'', (2) ``Unrelated'', and (3) ``Functional''. 
We label an \textsf{unrecognized} email as malicious if the content of the email is clearly spam or scam. 
We identify 76 malicious emails received on the honeytokens shared with 3 apps.
Some examples of malicious content include ransomware scam \cite{sextortionScamKrebOnSecurity} or Viagra spam \cite{viagraSpam}.
We label an \textsf{unrecognized} email as ``unrelated'' if the content of the email is not relevant to either the app or the app's host website. 
We identify 79 unrelated emails received on the honeytokens shared with 9 apps.
Some examples of unrelated content include promotional offers, links to product listings, and newsletters. 
Note that an unrelated email may be in violation of Facebook's TOS~\cite{tos} if the app has not clearly notified the user about data usage by other entities in their privacy policy. 
We label the content of an \textsf{unrecognized} email as ``functional'' if it is related to the core functionality of the app or the app's host website.
We identify 198 functional emails received on the honeytokens shared with 4 apps.

Figure \ref{fig: datamisuse} shows the \textsf{recognized} emails and \textsf{unrecognized} emails labeled as either malicious, unrelated, or functional received on the honeytokens shared with 16 Facebook apps from disclosed and unknown senders.
Next, we discuss the taxonomy of the unrecognized use of our shared honeytokens based on the disclosure test and the content analysis.

\textit{1) Malicious Content \& Unknown Relationship.}
This type of unrecognized use is the most egregious case of data misuse since the user data has been obtained by spammers or scammers who are sending malicious emails.
Note that we only receive malicious emails from unknown senders.
Specifically, we received 76 malicious emails on honeytokens shared with 3 apps from unknown senders indicated by red cross markers in Figure \ref{fig: datamisuse}.
Out of these 3 apps, the honeytokens shared with 2 apps ``Safexbikes Motorcycle Superstore'' and ``Printi BR API'' receive a ransomware scam email \cite{sextortionScamKrebOnSecurity}.
This email asks for ransom through bitcoin falsely claiming that the victim's browsing history has been compromised.
The honeytoken shared with the remaining one app ``WeWanted'' has received a large number of spam emails such as Viagra spam.
We also find anecdotal evidence of the potential breaches of the host websites of the two apps ``Safexbikes Motorcycle Superstore'' and ``Printi BR API''~\cite{safexbikesHackNotics,printibrbreach}.
We surmise that the attackers likely acquired our honeytoken due to either unintended exposure or breach of these app's server.
To date, we have not received any disclosure from any of these apps' host websites about a data breach.
We believe that \name can help detect such cases of data misuse where users' data shared with Facebook apps are obtained by spammers or scammers.

\textit{2) Unrelated Content \& Unknown Relationship.}
While the content is not malicious, this type of unrecognized use is worrisome for users since user data shared with Facebook apps have been potentially misused by an unknown entity.
We receive 9 unrelated emails on honeytokens shared with 4 apps from unknown senders indicated by blue plus markers in Figure \ref{fig: datamisuse}.
By analyzing the privacy policy of these 4 apps, we find that only 1 app (``NYX ca'') explicitly mentions the potential use of data by affiliate partners.
For the remaining 3 apps, these \textsf{unrecognized} emails are clearly data misuse since 1 app (``MyJapanBox'') does not have a privacy policy while other 2 apps (``PopJulia'', ``JustFashionNow'') do not explicitly mention potential use of user data by other entities. 
We believe that \name can help detect such cases of misuse where data shared with Facebook apps have been sold or transferred to unknown entities.

\begin{table}[!t]
\scriptsize
\centering
  \begin{tabular}{c|c|c|c}
    \textbf{App Name} &\textbf{Host}    & \textbf{Global} & \textbf{Country}\\
     \textbf{       } & \textbf{Website}   & \textbf{Alexa} & \textbf{Alexa}\\
      &     & \textbf{Rank} & \textbf{Rank}\\
    \hline
     Safexbikes Motor& safexbikes    & 89K & 8K (IN)\\ 
      cycle Superstore& .com      & & \\
    \hline
    WeWanted & wewanted.com.tw& 99K  & -\\ 
         \hline
     Printi BR API & printi.com.br& 15K & 441(BR)\\ 

\hline
     JustFashionNow &justfashionnow.com & 51K & 223 (MO)\\ 

    \hline
     PopJulia &popjulia.com & 469K & \\ 

   \hline
    MyJapanBox & myjapanbox.com  & 766K & - \\ 
    \hline
     Nyx CA & nyxcosmetics.ca   & 258K & 16K (CA) \\ 
     
    \hline
    Tom's Hardware &tomshardware    & 870 & 726 (USA) \\
    Guide-IT Pro  &.com    & &\\
    
    \hline
    Alex's first app  & beautymaker.com.sg    & 680K & 3K (SG) \\ 
    
    \hline
    Thailand Pro- & thailand-pr-  & 98K & 3K (TH) \\ 
    perty Login & operty.com   & &\\

    \hline
     Hop-On, & hop-on-hop   & 161K & 77K (USA)\\ 
    Hop-Off & -off-bus.com  & &\\

     \hline
    
     Leiturinha & leiturinha.com.br& 114K & 5K (BR)\\ 
\hline
    The Breast Expa & bestoryclub  & 484K & -\\ 
    nsion Story Club & .com   & &\\

    \hline
     Jacky's Electronics  & jackyselectronics & 517K & 8K (UAE)\\ 
      &   .com   & & \\
     \hline
     Berrykitchen.com & Berrykitchen.com  & 494K & -\\
    \hline
    uCoz.es Login & ucoz.es    & 157K & -\\ 
    &       & &\\

  \end{tabular}
  \caption{Popularity of 16 Facebook apps that are responsible for the unrecognized use of the shared honeytokens.} 
  \postcaption
\postcaption
\postcaption
\postcaption
\postcaption
\postcaption
\postcaption
\label{tbl: dataMisuseEmails}

\end{table}

\textit{3) Unrelated Content \& Disclosed Relationship.}
An unrelated email from a disclosed sender appears less worrisome for users as compared  to an unrelated email from an unknown sender.
Yet, we argue that our honeytokens shared with the Facebook apps have been potentially misused by the unrecognized senders to send emails not relevant to the app (which may also be in violation of Facebook's TOS~\cite{tos}).
We receive 70 unrelated emails on honeytokens shared with 5 apps from disclosed senders indicated by orange diamond markers in Figure \ref{fig: datamisuse}.
%
%

\textit{4) Functional Content \& Disclosed Relationship.}
This type of unrecognized use typically occurs when a Facebook app is using a partner website or an external service (e.g., a user authentication system) to send emails to their users.
We receive all 198 functional emails on honeytokens shared with 4 apps from disclosed senders indicated by grey square markers in Figure \ref{fig: datamisuse}.
Facebook's TOS~\cite{tos} permits the sharing of data with partner websites or external services if they use it to provide services within the app.
Hence, we do not consider these functional emails as a misuse of our honeytoken.
%

\presub \subsubsection{Potential Impact} \postsub
Table \ref{tbl: dataMisuseEmails} lists the 16 apps that are responsible for the unrecognized use of our honeytoken along with the global and country-level Alexa rankings of their host websites. 
While the global rank of their host websites ranges between 870 and 766K, half of them are ranked among the top 10K at the country-level.
Therefore, we argue that unrecognized use of data by these apps on popular websites potentially impacts a large number of users. 
For example, according to Alexa, the host website of ``Tom's Hardware Guide-IT Pro'' app has more than 10 million unique monthly visitors.

\presub 
\subsection{Misuse Detection Using Facebook's Ad Transparency Tool} 
\postsub 
\name also uses Facebook's ad transparency tool \cite{fbAdTransparencyHowTo} as the monitoring channel to detect potential misuse of data shared with Facebook apps.
Facebook's ad transparency tool enables \name to identify the apps whose shared honeytoken email addresses were uploaded to Facebook for ad targeting \cite{customAudienceAdsFB}.
To this end, we took 34 snapshots of the advertisers listed by Facebook's ad transparency tool over the period of one month.
We are able to identify 47 unique advertisers that uploaded our honeytoken email address for ad targeting.
Unfortunately, Facebook's ad transparency tool only provides the names of the advertisers but not the email address uploaded by the advertisers.
Thus, we cannot attribute the responsible apps since the Facebook account has been associated with multiple honeytoken email addresses. \footnote{We could generate a mapping between apps and advertisers by capturing a snapshot of advertisers after rotating an email address and installing a new app.
However, if we observe an \textsf{unrecognized} advertiser after installing a new app, it is not possible to distinguish whether the current app leaked our honeytoken or another previously installed app leaked it.}

While we cannot attribute the responsible Facebook app, we can still detect whether any of our shared honeytokens has been potentially misused by advertisers.
To this end, we match 47 advertisers' names and domains with each of the 1,024 apps using the process explained in Section \ref{ssec:detectabuse}.
Out of these 47 advertisers, 38 advertisers are \textsf{recognized} while the remaining 9 advertisers are \textsf{unrecognized}.
We further evaluate the accuracy of our labeling of advertisers as \textsf{recognized} or \textsf{unrecognized} through manual inspection. 
For a \textsf{recognized} advertiser,  we visit the Facebook page and web page of the advertiser and compare it against the matched app.
For an \textsf{unrecognized} advertiser, we visit the Facebook page and web page of the advertiser and then compare it against each Facebook app to check whether there is any potential relationship between an app and the advertiser.
Our manual inspection shows that we label advertisers as \textsf{recognized} and \textsf{unrecognized} without any mistakes. 
Overall, our results show that Facebook's ad transparency tool can also be leveraged to detect potential misuse of data shared with Facebook apps.
While the limited information provided by Facebook's ad transparency tool does not allow \name's current implementation to identify the Facebook apps responsible for the misuse, a modified implementation of \name can address this limitation.
Specifically, \name would be able to correctly attribute the app behind leaking our honeytoken to the advertiser if we created Facebook accounts on a per-app basis.
Note that we do not create a separate Facebook account for each of the 1,024 apps monitored in our array framework because of Facebook's stringent countermeasures against bulk account creation~\cite{fbFakeAccountsActions}.

\presub \subsection{Evaluation of Matrix Framework} \postsub  \label{ssec:evaluationmatrix}
Next, we analyze the matrix framework's effectiveness in attributing third-party apps to received emails and detecting data misuse. 
\name's matrix framework is deployed to share email honeytokens to the same 1,024 apps around the same time frame. 
Thus, we can use the array framework as ground truth to evaluate the accuracy of the matrix framework. 
Since the matrix framework shares an email honeytoken with multiple apps and two email honeytokens are shared with each app, we first need to attribute the received emails to the right apps. 
After attributing the responsible apps, we then use the same process as the array framework to label the received emails as \textsf{recognized} or \textsf{unrecognized}.

\noindent \textbf{Attribution.}
We receive a total of 23,303 emails on the email accounts associated with the two Facebook accounts $R$ and $C$ of the matrix framework.
Out of these 23,303 emails, we receive 11,370 emails on the email accounts associated with the Facebook account $R$ and 11,933 emails on the email accounts associated with the Facebook account $C$.
For attributing these 23,303 emails to 1,024 apps, using the process described in Section \ref{ssec:deployment}, we check whether the emails received on the email accounts of the two honeytokens shared with an app are from the same sender.
We are able to attribute 20,867 emails to 279 apps while 12 emails are excluded because they are conflicting.
Next, we analyze the remaining 2,424 unattributed emails.

To help understand the limitations of the matrix framework's attribution approach, we attempt to attribute the 2,424 unattributed emails using the following 3-step process. 
First, we leverage the array framework to generate a sender-to-app mapping between a Facebook app and the emails received on its associated email account. 
Note that this mapping is only available because the array and matrix frameworks are simultaneously deployed. 
Thus, we would not have this information for validation if the matrix framework was deployed standalone.
The sender-to-app mapping from the array framework allows us to attribute 1,972 emails to 59 apps.
Second, we use the keyword matching process detailed in Section \ref{ssec:detectabuse} to attribute the remaining 452 unattributed emails to apps.
We use the intuition that the keyword matching process labels an email as \textsf{recognized} if it is received from the app itself.
Hence, we attribute an email to an app if the email is labeled as \textsf{recognized} against the app.
This matching process allows us to attribute 282 emails to 25 apps.
Finally, we rely on manual eyeball analysis to attribute the remaining 170 emails to 3 apps. 
These apps are not attributed by the previous two steps because these apps either sent us \textsf{unrecognized} emails or they show non-deterministic behavior as discussed below. 
Overall, we are able to attribute the 2,424 unattributed emails to 82 unique apps using this 3-step process.

Next, we manually analyze our logs (including screenshots) collected during \name's automated workflow (shown in Figure \ref{fig: crawlerWorkflow}) to understand the root cause of unattributed emails in the matrix framework.
We identify two main reasons. 
First, we find that 25 apps are unattributed because only one of the two honeytokens associated with either Facebook account $R$ or Facebook account $C$ was successfully shared due to \textit{implementation issues}.
While our automation is generally robust, some unhandled errors occurred due to unexpected changes in Facebook's or the host site's web interface.
Second, we find that the 57 apps are unattributed due to \textit{non-deterministic app behavior} despite successfully sharing honeytokens with these apps.
Recall that the attribution requires that an app sends the same email on the email accounts of both honeytokens shared with the app and the sender's email address of both these emails are also the same.
We find that 50 out of these 57 apps remain unattributed because these apps fail to send emails on the email accounts of one of the two honeytokens shared with the app.
The remaining 7 apps are unattributed because these apps send the same emails using different email addresses on the two honeytokens shared with the app.
The unattributed apps due to the non-deterministic app behavior also indicate a limitation of the matrix framework to detect some cases of data misuse. 
As we also show later, the matrix framework is unable to detect misuse (i.e., false negatives) if an attacker sends emails using two different email addresses on the email accounts of both honeytokens shared with the app.


\noindent 
\textbf{Accuracy of misuse detection.}
Next, we use the process explained in Section \ref{ssec:detectabuse} to label the 20,358 emails as \textsf{recognized} and 509 emails as \textsf{unrecognized}.
The 509 emails labeled as \textsf{unrecognized} are attributed to 9 Facebook apps. 
We evaluate the matrix framework's data misuse detection accuracy using the array framework as ground truth because it does not have the issue of unattributed apps due to its one-to-one app-to-honeytoken mapping.
Using this ground truth, we define the following: (1) \textit{true positive} is an app labeled as \textsf{unrecognized} by both matrix and array frameworks; (2) \textit{true negative} is an app labeled as \textsf{recognized} by both matrix and array frameworks; (3) \textit{false negative} is an app labeled as \textsf{recognized} by the matrix framework but labeled as \textsf{unrecognized} by the array framework; and (4) \textit{false positive} is an app labeled as \textsf{unrecognized} by the matrix framework but labeled as \textsf{recognized} by the array framework.

For the 1,024 monitored apps, the matrix framework has 9 true positives and 1,008 true negatives.
Our results show that the matrix framework is able to detect most of the \textsf{unrecognized} apps while correctly labeling all of the apps as \textsf{recognized}.
Our matrix framework has no false positives and 7 false negatives.
Out of these 7 false negatives, one app is unattributed due to implementation issues, 5 apps are unattributed due to a non-deterministic behavior of these apps, and the remaining one app is unattributed due to conflicting emails discussed in Section \ref{ssec:deployment}.
A robust implementation of our automated workflow (Figure \ref{fig: crawlerWorkflow}) that handles more corner cases and unexpected errors can reduce unattributed apps. 
We can also apply ideas from error correction to the matrix framework by sharing honeytokens to apps in rows and columns (as shown in Figure \ref{fig: LinMat}) redundantly.  
This redundancy can help us reduce the likelihood of unattributed apps similar to prior literature on encoding messages using error correction codes \cite{HammingECC}. 

Overall, our results highlight a trade-off between accuracy and scalability --
the matrix framework is less accurate but more scalable than the array framework if email addresses rotation is a bottleneck.

\presec \section{Discussions \& Limitations}

\subsection{Data Deletion by Third-party Apps}  \label{ssec:datadeletionrequest} \postsub
Facebook users have a limited recourse of action to exercise control over the data that is retrieved by third-party apps. 
Facebook provides users the option to uninstall the app but this will only prevent the app from accessing user data in the future.
Note that user data already retrieved by an app, before app uninstallation, may still be stored on the app's server. 
After uninstalling an app, the user can request the app to delete their data \cite{fbAppSettingsTab}.
Facebook recommends app developers to implement a one-click data deletion request callback \cite{fbDataDelReq}.
Unfortunately, third-party apps currently do not implement the recommended data deletion request callback.
As an alternate, Facebook suggests users to contact the app developer on their own by typically pointing them to the app's privacy policy. 
Facebook's TOS require app developers to honor their users' data deletion requests \cite{tos}.
Next, we investigate the process of requesting third-party Facebook apps to delete user data and also evaluate its effectiveness.

\noindent \textbf{Contacting app developers.} 
To evaluate the effectiveness of the data deletion request process, we conduct a follow-up experiment where we contact third-party app developers and ask them to delete our data (including honeytoken email addresses) shared with these apps. 
To this end, we sample 100 apps out of the 332 apps installed in the array framework that sent us at least one email.
Note that sampling from these 332 apps ensures that our data has been stored by these apps since we receive one or more emails from them.
We gather the contact URLs provided by these apps to users upon installation. 
We then open the contact URLs provided by these apps and manually analyze them to identify developer contact information, which is typically a contact email or contact form. 
Out of the 100 apps, we find that 87 apps provide a contact email while the remaining 13 apps provide a contact form. 
We then contact the app developers to delete the data retrieved from our Facebook account, including the honeytoken email address. 
Specifically, we email the data deletion requests to the email address provided by 87 apps and submit the data deletion requests through the contact forms of the remaining 13 apps.
Out of the 87 contact emails, our email server is able to successfully send emails on 80 while the remaining 7 emails failed due to delivery errors. 
Out of the 13 contact forms, we are able to successfully submit 7 contact forms, while the submission of the remaining 6 contact forms failed due to website errors.
In total, we are able to successfully contact 87 apps through the contact email or the contact form.



\begin{table}
\scriptsize
\centering
  \begin{tabular}{c|c}
  
Requested data deletion successfully & 87 apps \\
   \hline
  Not responded & 42 apps \\
   \hline
  Responded & 45 apps \\
   \hline
   Acknowledged data deletion & 29 apps \\
   \hline
   Send email(s) after acknowledging data deletion & 13 apps \\
  \end{tabular}

  \caption{Our results show that the process of contacting Facebook apps to request data deletion is mostly ineffective. Approximately only half of the Facebook apps responded back to our request. 13 Facebook apps even send us one or more emails despite acknowledging that they deleted our data.}
\label{tbl: contactDevs}
\postcaption
\postcaption
\postcaption
\postcaption
\postcaption
\postcaption
\postcaption
\end{table}

\noindent \textbf{Response by app developers.}
We contacted these 100 app developers in April 2019 and waited over a month to give them sufficient time to respond to our data deletion requests. 
Table \ref{tbl: contactDevs} summarizes our results.
Only 52\% of the apps (45 out of 87) responded back to our requests while the remaining 48\% of apps (42 out of 87) did not respond.
We manually analyze these responses to check whether or not these apps confirm data deletion.
Out of these 45 apps that responded, only 29 acknowledged that they have deleted our data (or canceled our user account).
The remaining 16 apps either forwarded our requests to relevant persons in their respective organizations, request additional steps on our part such as removing the account from the app's associated host website, or deny storing any data retrieved through the Facebook app.

We also continue to monitor emails received from these 87 apps to check whether they keep using our data despite our deletion request.
We receive at least one email from 49 out of these 87 apps after submission of our data deletion request. 
It is noteworthy that 13 of these 49 apps earlier confirmed that they have deleted our data.
Our analysis of contacting developers to request data deletion raises two key issues.

First, we note that the process to request data deletion is hard to navigate for a lay user.
Facebook currently does not play any active part in the data deletion process. 
Facebook completely relies on third-party app developers to fulfill users' data deletion requests.
Users have to go through a cumbersome process of finding ways to contact app developers, which typically entails reading a long privacy policy on the app's host website to find the contact information. 
Even when a user is able to successfully contact the app developer despite these challenges, our results show that many apps require significant back-and-forth communication which further complicates this process for a lay user. 
We argue that Facebook should (at the very least) mandate the developers to implement data deletion request callback \cite{fbDataDelReq} into their apps.
This would not only provide a user-friendly mechanism for requesting data deletion but also help Facebook audit compliance of Facebook's TOS by third-party apps.

Second, we note that the process of contacting Facebook apps is mostly ineffective. 
Our results show that approximately half of the Facebook apps respond back to our data deletion requests. 
Out of the apps that respond to our requests, less than two thirds of them acknowledge that they have deleted user data.
Even when the apps acknowledge that they have deleted user data, there is no way to guarantee that our data has been actually removed.
In fact, we continue to receive emails from 13 apps that earlier confirmed our data has been deleted.
We believe that \name can help Facebook detect such cases.

\presub 
\subsection{Lack of Enforcement by Facebook} 
\postsub
We observe several instances of lack of enforcement of existing Facebook's developer policies.
Facebook's TOS require third-party apps to provide a privacy policy that explains the collection and use of data \cite{tos}. 
Out of the 1,024 Facebook apps analyzed in our experiments, 6\% apps (62 out of 1,024) do not provide a privacy policy.
For the remaining 94\% apps that do have a privacy policy, we note that many apps use cookie-cutter policies that do not comply with Facebook's TOS.
For instance, Facebook requires an app's privacy policy to clearly explain how users can request the deletion of their data. 
We find that the privacy policies of many apps do not provide this information. 
We argue that Facebook should more strictly enforce these policies. 
%
It is noteworthy that even when apps provide a compliant privacy policy, Facebook does not have a sound mechanism to check whether the apps are actually in compliance.  
Facebook currently relies on certification of TOS compliance by third-party app developers as well as a mandatory compliance audit of the app's infrastructure by Facebook or an independent auditor \cite{tos}. 
However, neither Facebook can actually verify the authenticity of these certifications of compliance (as shown by the Cambridge Analytica episode \cite{CambridgeAnalyticaFbNewsRoom}) nor Facebook can mandate a compliance audit (e.g.,  \cite{rankwaveFacebook}) to detect potential violations.
We believe that \name can help Facebook detect such violations of its TOS.

\presub \subsection{Limitations and Future Extensions} \postsub 

\descr{Random emails.} 
Due to the low cost of mass email spam, spammers send emails to  email servers randomly by creating arbitrary email addresses \cite{pitsillidisJudo10NDSS,StoneGrossEmailSpamEconomy11Leet}. 
It is possible that a spammer may end up sending an email to one of the honeytokens used by \name. 
To avoid incorrect misuse detection in this case, our email server tries to exclude such emails by using a catch-all email account \cite{catchallemailAcctGodaddy}.
Specifically, our catch-all email account receives all emails sent to incorrect email addresses on the email server that do not exist.
This catch-all email account acts as a control account to eliminate such random emails from spammers.
Note that our catch-all email account did not receive any spam  during our experiments.

\descr{Use of other email service providers.} 
Using our custom domain for creating email accounts may result in false negatives.
First, it is possible that the attacker may not be interested in our honeytoken (that uses our custom domain) due to its lack of established reputation. 
Second, an attacker aware of \name's deployment may try to detect and exclude the honeytoken email addresses of domains that are known to deploy \name. 
We argue that partnerships with popular email providers can address these issues.

\descr{Other honeytokens and monitoring channels.} 
We acknowledge that our  implementation of \name would not detect misuse of data shared with third-party apps if their behavior is not observable through the honeytoken and monitoring channels that are currently used. 
\name's current implementation is limited to email address as the honeytoken and two monitoring channels (received emails and Facebook's ad transparency tool).
Future improvements to \name should explore using other honeytokens and monitoring channels. 
First, \name can be extended to use other profile attributes such as location, gender, and date of birth as honeytokens.
However, it is challenging to detect the misuse of these attributes due to a lack of sound monitoring channels.
Second, \name can be extended to use other monitoring channels to detect misuse of honeytokens.
For example, \name's existing monitoring channels cannot detect the misuse when an attacker uses a honeytoken as an identifier to link with other data sources. 
One example of such an attacker could be data brokers who typically gather information on unique identifiers including email addresses from various sources such as trackers, advertisers, or other partners.
Future work can consider periodically querying data broker APIs~\cite{acxionAPI,experianAPI} as a monitoring channel.
We could not use data brokers in our experiments because most of them have proprietary APIs that are not openly available to researchers.

\descr{Scaling \name.}
We can reduce the number of required honeytokens by using higher-dimensional tensor frameworks. 
As discussed earlier, a 1-dimensional (array) framework requires $N$ honeytokens while a 2-dimensional (matrix) framework requires $2\times\sqrt[2]{N}$ honeytokens to monitor $N$ apps.
Likewise, an $n$-dimensional framework would require only $n\times\sqrt[n]{N}$ honeytokens to monitor $N$ apps. 
Thus, as we increase dimensions from 1 to 2 to monitor $N=1,000,000$ apps, we can reduce honeytokens from $N=1,000,000$ to $2\times\sqrt[2]{N} = 2,000$.
However, there is a trade-off between scalability (i.e., reducing the number of honeytokens) and accuracy (i.e., false negatives due to the non-deterministic app behavior discussed in Section \ref{ssec:evaluationmatrix}).

We denote the probability that a honeytoken shared with an app does not receive a \textit{target} email due to the non-deterministic behavior as $P(ht_i)$.
Let's assume an arbitrary probability $P(ht_i)=\epsilon$.
In the case of a 1-dimensional framework, we share one honeytoken with an app.
Hence, the probability that we do not receive the target email on the shared honeytoken due to the non-deterministic behavior will be $\epsilon$.
In the case of an $n$-dimensional framework, we share exactly $n$ honeytokens with an app.
Assuming independence, the probability that we do not receive an email on one of the $n$ honeytokens will be $[P(ht_1) \vee P(ht_2) \vee ... \vee P(ht_n)] = 1-(1-\epsilon)^n$.
Since we cannot attribute the target email to its app if any of the $n$ honeytokens do not receive the email, the probability of unattributed apps using an $n$-dimensional framework will be $1-(1-\epsilon)^n$.
Thus, the probability of unattributed apps increases at higher dimensions.

%

\descr{Data misuse on other platforms.} 
\name can be adapted to detect misuse of data shared with third-party apps on other social networking platforms such as Twitter \cite{TwitterapiDev}, Instagram \cite{InstagramApiDev}, and Snapchat \cite{SnapChatapiDev} as well as other online platforms that support third-party apps~\cite{GmailapiDev,GsuiteapiDev}.
%
It is noteworthy that the underlying methodology of \name remains the same to monitor apps on other platforms. 
We only need to modify some parts of \name's existing implementation while we can still reuse various parts. 
For example, \name can be adapted to monitor third-party apps on Twitter by modifying the existing implementation to associate a honeytoken to a Twitter account instead of a Facebook account and the automation of host websites to install twitter apps instead of Facebook apps. 
%
We can reuse the existing infrastructure to monitor apps once the honeytoken has been shared with a third-party app.
For example, we can continue using our email server to monitor data misuse through received emails. 
Nonetheless, the feasibility of reusing a monitoring channel may vary. 
For example, the ad transparency tool is available on Facebook but not on Twitter.
%


\presec 
\section{Conclusion} 
\postsec
Third-party apps on online social networks with access to users' personal information pose a serious privacy threat.
A slew of recent high-profile scandals demonstrate that third-party apps are being exploited to harvest and misuse data of millions of users. 
We presented \name to help independent watchdogs detect misuse of data shared with third-party apps without needing cooperation from
online social networks.
Our deployment of \name uncovered several cases of misuse of data shared with third-party apps on Facebook including ransomware, spam, and targeted advertising. 
Our results also demonstrated that Facebook does not fully enforce its TOS as many apps do not disclose their data  sharing practices or honor data deletion requests by users.
A larger-scale longitudinal deployment of \name can potentially uncover previously undetected cases of misuse of data shared with third-party apps on Facebook as well as other online social networking platforms.

Last but not least, we discuss the ethical considerations associated with our study. 
We did not seek the consent from Facebook or the third-party apps monitored in our study because it could potentially threaten the validity of our findings. 
The small number of Facebook accounts and automated crawlers used in our study posed only modest overheads for Facebook and third-party app developers. 
We believe that any potential harms from our study are outweighed by the concrete benefits to Facebook and the general public. 
To foster follow up research on privacy issues of third-party apps on online social networks, the code and data of \name is available at \texttt{github.com/shehrozef/CanaryTrap/}.

\presec 
\section*{Acknowledgments} 
\postsec \postsec 
We would like to thank Kathy Zhong, Secondary Student Training Program (SSTP) scholar from Amador Valley High School, for her help with theoretical analysis of CanaryTrap. 
This work is supported in part by the National Science Foundation (grants 1715152, 1815131, and 1954224) and a grant from the Higher Education Commission (HEC) Pakistan.






\bibliographystyle{abbrv}
\bibliography{main}
\vfill\eject

\end{document}